\documentstyle[multicol,aps,prl,epsf]{revtex}
\begin{document}

\draft
\title{One-dimensional ordered structure of $\alpha$-sexithienyl on Cu(110)}

\author{Manabu Kiguchi, Shiro Entani, Genki Yoshikawa, Koichiro Saiki}
\address{Department of Complexity Science $\&$ Engineering, Graduate
School of Frontier Sciences, The University of Tokyo, \\7-3-1 Hongo,
Bunkyo-ku,Tokyo 113-0033, Japan}

\date{\today}

\maketitle

\begin{abstract}

We have studied atomic structures of $\alpha$-sexithienyl (6T) films grown on Cu(110) by near-edge x-ray absorption fine structure (NEXAFS). A one-dimensional (1D) ordered structure of 6T with its molecular long axis parallel to the Cu[001] direction could be fabricated by deposition at 300 K and subsequent annealing at 360 K. Polarization and azimuth-dependent NEXAFS revealed the formation process of the 1D structure and showed the molecular orientation in the in-plane direction directly. We propose here a method to obtain the orientation distribution function of molecules using NEXAFS.

\end{abstract}

\medskip

\begin{multicols}{2}
\narrowtext

Atomic and electronic structures of organic/metal interfaces have attracted wide attention for versatile properties and technological applications. Currently, researches are not limited in the structural analysis, but are extended to intentional structural control aiming at fabrication of nano structures. For example, the molecular orientation of $\alpha$-sexithienyl (6T) grown on Cu(111) can be controlled by changing the substrate temperature in the out-of-plane direction (perpendicular to the surface)\cite{1}. With respect to the orientation control in the in-plane direction (parallel to the surface), a one-dimensional (1D) structure of pentacene was fabricated on Cu(110) by deposition at 300 K and subsequent annealing at 400 K \cite{2}.

In the present study, we tried to control molecular orientation of 6T on Cu(110) in both out-of-plane and in-plane direction. 6T is one of the most promising ƒÎ-conjugated organic materials due to its fascinating properties (eg. high FET mobility\cite{3}) and the ability to tune its physical properties by changing chain length or substituent groups. The structure of the 6T film was studied by near-edge x-ray absorption fine structure (NEXAFS). In contrast to diffraction methods, which can only probe the ordered region in the specimen, NEXAFS can determine the average structure of the sample. Therefore, we could study the formation mechanism of the ordered structure using NEXAFS. Furthermore, we could directly determine the molecular orientation in the in-plane direction by azimuth-dependent NEXAFS, and orientation distribution function of 6T. 

Mechanically and electrochemically polished Cu(110) was cleaned by repeated cycles of Ar$^{+}$ sputtering and annealing. Molecules of 6T were evaporated from a Knudsen cell. Measurements of S K-edge NEXAFS were carried out at the soft x-ray double-crystal monochromator station BL-11B of the Photon Factory in the Institute of Materials Structure Science\cite{4}. The fluorescence yield detection method was adopted to obtain S K-edge NEXAFS.

\begin{figure}
\begin{center}
\leavevmode\epsfysize=50mm \epsfbox{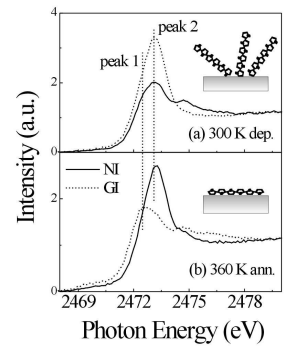}
\caption{S K-edge NEXAFS spectra of the 4 \AA {} thick 6T film grown on Cu(110); (a) after deposition at 300 K, and (b) subsequent annealing at 360 K. The line and dot-line correspond to spectra at the x-ray angle of 90$^{\circ}$(NI) and 30$^{\circ}$(GI). In the inset, the structural models of the 6T film.}
\label{fig1}
\end{center}
\end{figure}

Figure~\ref{fig1} shows the polarization-dependent NEXAFS of the 4 \AA {} thick 6T film grown at 300 K (Fig.~\ref{fig1}-a) and subsequent annealing at 360 K (Fig.~\ref{fig1}-b). Two intense peaks are observed at 2472.5 eV (peak 1) and 2473.1 eV (peak 2), and these peaks show noticeable polarization dependence. Intensity of peak 2 is larger at the grazing x-ray incidence (30$^{\circ}$) in Fig.~\ref{fig1}-a, while in Fig.~\ref{fig1}-b it becomes larger at the normal x-ray incidence, instead peak 1 is clearly observed at grazing x-ray incidence. From the discussion in the previous study\cite{1}, peak 1 is assigned to the S1s-to-$\pi ^{\ast}$ transition with its transition moment is normal to the ring face, while peak 2 to the S1s-to-$\sigma ^{\ast}$ (S-C) transition with its moment is parallel to the molecular long axis. Based on this peak assignment, the mean inclination angle of the 6T molecules from the surface normal is determined to be 50($\pm$10) $^{\circ}$ for the film grown at 300 K (Fig.~\ref{fig1}-a), and 10($\pm$10) $^{\circ}$ for the film annealed at 360 K. The value of 50$^{\circ}$ suggests that 6T molecules randomly adsorbed on Cu(110) at 300 K.

\begin{figure}
\begin{center}
\leavevmode\epsfysize=50mm \epsfbox{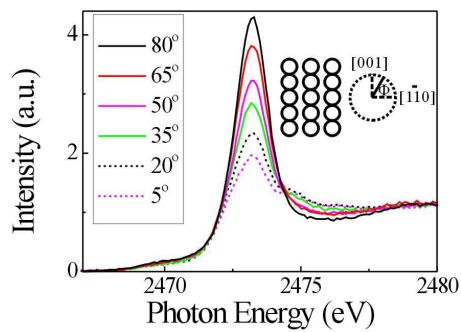}
\caption{The azimuth-dependent NEXAFS of 4 \AA {} thick 6T/Cu(110) taken at normal x-ray incidence. The Cu[1$\overline{1}$0] and [001] direction are defined as 0$^{\circ}$ and 90$^{\circ}$.}
\label{fig2}
\end{center}
\end{figure}

\begin{figure}
\begin{center}
\leavevmode\epsfysize=50mm \epsfbox{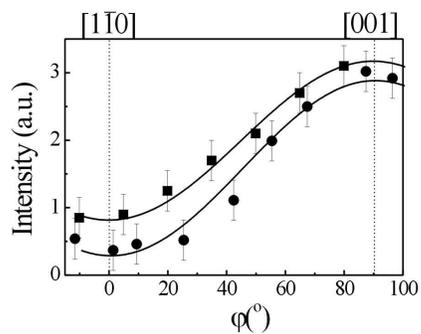}
\caption{The intensity of $\sigma ^{\ast}$ peak as a function of the azimuthal angle for the 4 \AA {} thick 6T film grown on Cu(110) (boxes) and Ag(110) (circles).}
\label{fig3}
\end{center}
\end{figure}

We, then, studied the molecular orientation in the in-plane direction by azimuth-dependent NEXAFS.\cite{5} Figure~\ref{fig2} shows the azimuth-dependent NEXAFS taken at normal x-ray incidence for 6T/Cu(110). In the figure, 0$^{\circ}$ and 90$^{\circ}$ correspond to the Cu[1$\overline{1}$0] and [001] direction. The intensity of $\sigma ^{\ast}$ peak in the figure shows a clear azimuth dependence. Figure~\ref{fig3}  shows the intensity of $\sigma ^{\ast}$ peak determined by curve fitting analysis\cite{6}, as a function of azimuthal angle for 6T/Cu(110). The intensity of $\sigma ^{\ast}$ peak is largest at 90$^{\circ}$ and weakest at 0$^{\circ}$, showing that the molecular long axis is parallel to the Cu[001] direction. The above polarization and azimuth-dependent NEXAFS results indicate that we could fabricate a 1D ordered structure of 6T on Cu(110). The present NEXAFS results clearly show the formation process of the ordered structure; 6T molecules adsorbed randomly on Cu(110) at 300 K, and a thermally stable 1D ordered structure appeared by annealing. These NEXAFS results agree with the previous results of reflection high energy electron diffraction\cite{7}, in which clear diffraction pattern was observed only after annealing at 360 K. 

Here, it should be noticed that we could directly find the molecular orientation of the sample with twofold symmetry, by measuring NEXAFS as a function of azimuthal angle. In most of previous studies, NEXAFS were measured at only several points in the in-plane direction, and the molecular orientation was not directly shown. On the other hand, the molecular orientation is determined only after fitting the polarization-dependent NEXAFS to the theoretical curves in the out-of-plane direction\cite{6}.

\begin{figure}
\begin{center}
\leavevmode\epsfysize=50mm \epsfbox{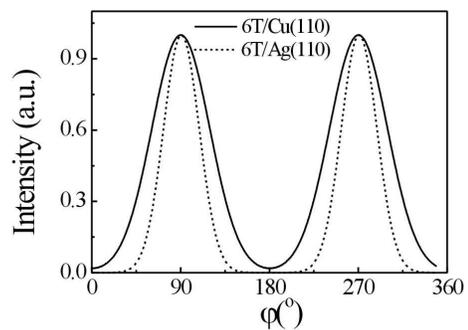}
\caption{The orientation distribution function of 6T molecules grown on Cu(110) and Ag(110).}
\label{fig4}
\end{center}
\end{figure}

Figure~\ref{fig3} also shows the result of 6T/Ag(110), in which a similar 1D structure of 6T is formed like 6T/Cu(110)\cite{5}. Although the azimuth dependence of $\sigma ^{\ast}$ peak intensity shows a similar tendency for both 6T/Cu(110) and 6T/Ag(110), they are quantitatively different from each other. We would, thus, introduce an orientation distribution function of molecules with twofold symmetry for clarifying the interface structure between 6T/Cu(110) and 6T/Ag(100). Here, the orientation distribution function is approximated by the following gauss function , 
\begin{equation}
f(\alpha)=\frac{1}{\sqrt{2 \pi \sigma}}exp(-\frac {1}{2 \sigma ^{2}}{\alpha ^{2}})
\end{equation}

where $\alpha$ and $\sigma ^{2}$ are the azimuthal angle between the molecular long axis and the metal substrate [001] direction, and variance. The intensity of the NEXAFS peak is proportional to $cos ^{2} \theta$, where $\theta$ is the angle between the electric vector of the x-ray and the transition moment of the peak. The intensity of $\sigma ^{\ast}$ peak observed by NEXAFS is, thus, represented as , 
\begin{equation}
I(\varphi)=\int_0^{2 \pi} f(\alpha)cos^{2} (\varphi - \alpha)d\alpha
\end{equation}

where $\varphi$ is the azimuthal angle between the electric vector of incident x-ray and [1$\overline{1}$0] direction. By fitting the experimental result with the fitting function, $\sigma$ was determined to be 29$^{\circ}$ and 18$^{\circ}$ for 6T/Cu(110) and 6T/Ag(110). $\sigma$ is smaller for 6T/Ag(110), that is, a better ordered 1D structure of 6T is formed on Ag(110). Figure~\ref{fig4} showed the obtained orientation distribution function ($f(\alpha)$) of 6T on Cu(110) and Ag(110).

Finally, we would discuss the formation mechanism of the 1D structure of 6T on Cu(110) and Ag(110). The interval of five-membered rings is 3.92 \AA {} for a 6T molecule\cite{8}. On the other hand, the Cu-Cu distance is 3.62 \AA {} in the Cu[001] direction and misfit between 6T and Cu is -7.7$\%$, while the Ag-Ag distance is 4.09 \AA {} in the Ag[001] direction and misfit is 4.3$\%$. The misfit between the interval of five-membered rings and Cu-Cu or Ag-Ag distance in other direction is larger than that in the [001] direction. Therefore, the 6T molecules grew on Cu(110) and Ag(110) with their molecular long axes parallel to the [001] direction. Since the misfit for 6T/Cu(110) is larger than that for 6T/Ag(110), more defects would be formed in the grown 6T film on Cu(110), leading to larger $\sigma$ of the orientation distribution function of the 6T molecules.  

In conclusion, a one-dimensional (1D) ordered structure of 6T with its molecular long axis parallel to the Cu[001] direction, could be fabricated by deposition at 300 K and subsequent annealing at 360 K. The polarization and azimuth-dependent NEXAFS study could reveal the formation process of the 1D structure and the orientation distribution function of 6T. This work was supported by a Grant-in-Aid for Creative Scientific Research, No. 14GS0207, MEXT and PF-PAC (2002G260).

\end{multicols}
\end{document}